\newcommand{\la}{\langle}
\newcommand{\ra}{\rangle}

\newcommand{\beq}{\begin{eqnarray}}
\newcommand{\eeq}{\end{eqnarray}}

\newcommand{\btem}{\bibitem}
\newcommand{\TK}{T.\ Kunihiro}

\newcommand{\PTP}{Prog.\ Theor.\ Phys.}
\newcommand{\MPR}{Phys.\ Rev.}

\documentclass[aps,prl,twocolumn,groupedaddress]{revtex4}
\usepackage{graphicx,epsf}
\usepackage[usenames]{color}

\begin{document}


\title{
Chiral Transition and Some Issues on the Scalar Mesons
}


\author{Teiji Kunihiro}
\affiliation{Yukawa Institute for Theoretical Physics, Kyoto University, Sakyoku,
Kyoto 606-8502, Japan}


\date{\today}

\begin{abstract}
Some issues on the low-mass scalar mesons are discussed in relation to 
the chiral transition of QCD vacuum.
The importance to explore  the possible collective  nature of the 
$\sigma$ meson is 
emphasized in association with the chiral properties of nuclear media.
\end{abstract}


\maketitle

\section{Introduction}

There are serious controversies on the nature
of the low-lying scalar mesons with a mass lower than 1 GeV, 
especially the $\sigma$ meson.
 In the non-relativistic constituent quark model,
 $J^{PC}=0^{++}$ is realized as a $^3$P$_0$ state, which 
implies that the mass of the $\sigma$ should be
in the 1.2-1.6 GeV region.
Therefore some mechanism is needed to down the mass.
The possible mechanisms so far proposed include:\\ 
(1)~ The color magnetic interaction 
between the di-quarks as advocated by Jaffe\cite{jaffe}.\\
(2)~ The collectiveness of the scalar mode as the pseudoscalar mode; 
a superposition of $q\bar q$ states, which collectiveness is
due to chiral symmetry\cite{njl}.\\ 
(3)~ The $\sigma$ meson may be a molecular states of
the NG bosons naively suggested by the fact that the
unitarized chiral dynamics could account for the 
existence of the $\sigma$ pole\cite{OOR}.\\ 
In this talk, I shall give some arguments to adovocate
 the viewpoint (2) above.

The basic idea  underlying my talk  is
 that  the low-energy hadron physics
may be regarded as  a study of the nature of QCD vacuum and
 hopefully its symmetry properties: In other words,
hadron physics is a condensed matter physics
of the QCD vacuum. In fact,
 the  QCD vacuum is realized non-perturbatively
and  hadrons are elementary excitations on top of the non-perturbative
vacuum, while QCD itself is  written solely in terms of
 quark and gluon fields.
Moreover, symmetries in (classical) QCD Lagrangian are not manifest;
color SU(3) is not manifest owing to the
confinement and (approximate) chiral symmetry is spontaneously broken.

Such a viewpoint as described above on the vacuum in quantum field theories  
was introduced by  Nambu\cite{nambu1960,njl}.
In the paper\cite{nambusuper} entitled ``Quasiparticles and 
Gauge invariance in the Theory of Superconductivity'',
Nambu showed that 
the appearance of the would-be massless collective mode 
in the broken phase is a logical consequence of the 
gauge invariance of the theory; here we remark that
the appearance of a massless mode which couples
to the longitudinal part of the electro-magnetic current
and hence insures the gauge invariant explanation of 
Meissner effect had been shown earlier by Bogoliubov\cite{bogoliubov} and
Anderson\cite{anderson-coll} and others\cite{others}.
Then in an important but relatively less known paper\cite{nambu1960} entitled
`` Axial Vector Current Conservation in Weak Interactions'',
he suggested that the fundamental theory of the
hadron dynamics could be obtained by
replacing the gauge invariance, the energy gap, and the
collective excitation in the theory of superconductivity 
with $\gamma_5$ invariance (chiral symmetry), baryon mass $m_f$,
and the pion: The pion automatically emerges as a bound
state of baryon pairs; the nonzero meson masses would indicate that 
the chiral symmetry is not rigorous.
This scenario was shown to be the case in a model 
calculation by Nambu and Jona-Lasinio \cite{njl}
in the famous paper entitled
``Dynamical Model of Elementary Particles Based on an Analogy with
Superconductivity'':
It is noteworthy here  
that a isoscalar-scalar meson with $J^{PC}=0^{++}$
emerges with the mass $2m_f$  as another 
collective mode as the pion does; the scalar meson is called the
$\sigma$ meson, which is named $f_0$ by the PDG.

Now the $\sigma$ meson has another important aspect.
First notice that the chiral transition is  a phase transition of 
 QCD vacuum with 
 $\la \bar{q}q\ra$ being the order parameter, as clearly shown
by recent lattice simulation of QCD\cite{karsch00,karsch02}.

If a phase transition is of 2nd order or {\em weak} 1st order,
there exists  ``soft'' modes which decreases it mass when the
system approach the critical point;
the soft modes are actually  fluctuations of the order parameter
of the phase transition\cite{soft}.
 For chiral transition,
the relevant fluctuation is described by $\la(:\bar{q}q:)^2\ra$,
which has the same quantum numbers as 
the $\sigma$-meson does, i.e., $(I=0, J^{PC}=0^{++})$.
Accordingly the $\sigma$ meson can become
 the soft mode of chiral transition at $T\not=0$ and/or
$\rho_B\not=0$\cite{ptp85}:
 $m_{\sigma}\rightarrow 0$, $\Gamma_{\sigma}\rightarrow 0$
\begin{figure}
\includegraphics[scale=.4]{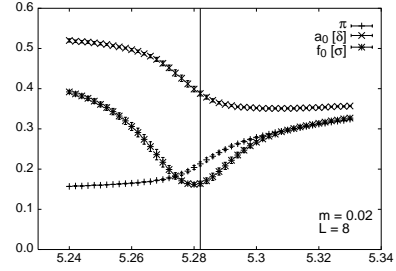}
\caption{
The lattice calculation \cite{karsch02} of the temperature dependence of 
the generalized masses as defined by $m_{\sigma}^2=\chi_{\sigma}^{-1}$
for the $\sigma$ meson
where $\quad \chi_{\sigma}=\la(\bar q q)^2\ra.$
\label{genmass}}
\end{figure}
A lattice calculation of the  generalized masses
\cite{karsch02} shows a behavior consistent with the soft mode
nature of the $\sigma$ meson, as shown in Fig.1:
One sees
(1)~ the softening of $\sigma$
(2)~ a degeneracy of the $\sigma$ and $\pi$ at high $T$.
Moreover one may notice that 
$U_A(1)$ symmetry is not restored even at high $T$;
$m_{\delta}\, \leftrightarrow\, m_{\pi}$.

But, what is
 the significance of the $\sigma$ in hadron physics\cite{kuni95,paris}? \\
a)~ A pole in the $\sigma$ channel 
 in the low-mass range is observed in the 
$\pi$-$\pi$ scattering matrix\cite{pipi,CLOSETORN}.
As a compilation of the pole positions of the $\sigma$ 
obtained in the modern 
 analyses, see \cite{summary}.
It is recognized that respecting  chiral symmetry,
unitarity and crossing symmetry is essential to reproduce the phase shifts
 both in the $\sigma\, (s)$- and $\rho\, (t)$-channels
with a low-mass $\sigma$ pole\cite{t-chan,igi}.\\
b)~ Moreover, the $\sigma$ is also 
seen in decay processes from heavy particles;
$D^{+}\rightarrow \pi^{-}\pi^{+}\pi^{+}$
\cite{aitala}.\\ 
c)~ As is well known\cite{obe}, 
the $\sigma$ with a mass $400\sim 600$ MeV is 
responsible for the intermediate range attraction in
the nuclear force; without the $\sigma$ contribution, any nucleus
can not be bound, nor possible our existence.\\
d)~ The $\sigma$ can accounts for $\Delta I=1/2$ enhancement in 
K$^0\rightarrow 2\pi$ 
compared with K$^{+}\rightarrow \pi^{+}\pi^{-}$\cite{mls}.\\ 
e)~  The empiricalvalue of the $\pi$-N sigma term 
$\Sigma_{\pi N}\sim$ 40-50 MeV may be accounted for 
 by the collectiveness of the $\sigma$
\cite{hk90}, as will be discussed shortly.

The quark content in the scalar channel of 
a nucleon, $\la N\vert \bar{q} q\vert N\ra$ 
can be enhanced by 
the collective $\sigma$ mode so much as to
 reproduce the empirical value of the 
$\pi$-N sigma term\cite{hk90}.
In fact,  using the quark contents given in Table 1, we have
\beq
\Sigma_{\pi N}&=&\hat{m}\la \bar{u}u+\bar{d}d\ra_{N} \nonumber \\ 
 &=& 5.5{\rm
MeV}\times
 (4.97+4)\simeq  50 \, {\rm MeV},
\eeq
which is similar to the empirical value in contrast to the value
given by  the naive quark model
$ 5.5\times (2+1)\simeq 17$  MeV.

\begin{figure}
\includegraphics[scale=.3]{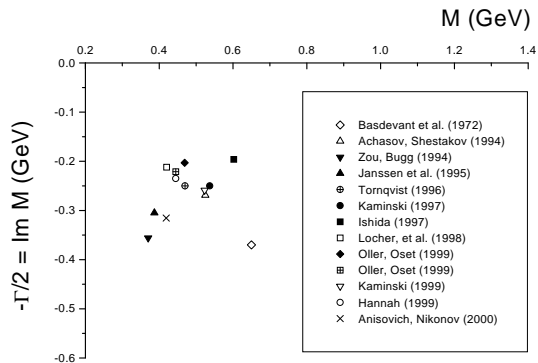}
\caption{
The poles of the $S$-matrix in the complex mass plane (GeV) for the 
$\sigma$ meson; complied in
\cite{summary}.
\label{compile}
}
\end{figure}
\begin{figure}
\rotatebox{-90}
{\includegraphics[scale=.3]{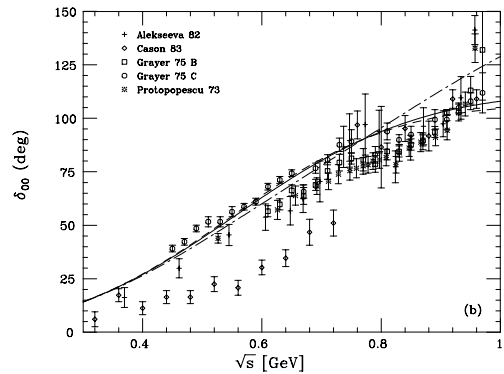}
}
\caption{The $\pi$-$\pi$ phase shift in the $\sigma$ channel 
calculated by means of $N/D$ method 
\cite{igi} reproduces the experimental data with the incorporation of 
the $\sigma$ pole in the $s$ channel as well as the 
$\rho$ pole in the $t$ channel; the phase shift obtained without
 the $\sigma$ pole fails in reproducing the empirical data.
\label{igi2}
}
\end{figure}


\begin{center}
\begin{table}
\caption{
Light quark contents of baryons calculated with a chiral
quark model with the $\sigma$ meson cloud incorporated
\cite{hk90}. 
The numbers in \textcolor{blue}{(\, )}
 are those in the naive quark model.
}
\begin{tabular}{|l|c|c|c|}
   \hline
\ \  B    & $\langle \bar uu\rangle_B$ & $\langle \bar dd\rangle_B$
& $\langle \bar ss\rangle_B$
\\ \hline \hline
P (938) & 4.97 \textcolor{blue}{(2)} & 4.00 \textcolor{blue}{(1)} 
& 0.53 \textcolor{blue}{(0)} \\
$\Lambda ^{0}$ (1115) & 3.63\textcolor{blue}{(1)}
 & 3.63\textcolor{blue}{(1)} & 1.74\textcolor{blue}{(1)} \\
 $\Delta ^{++}$ (1232) & 3.66\textcolor{blue}{(2)} & 0.76
\textcolor{blue}{(0)} 
& 0.26 \textcolor{blue}{(0)}
\\
$\Omega ^{-}$ (1672) & 0.72 \textcolor{blue}{(0)} & 0.72
\textcolor{blue}{(0)} 
& 3.71 \textcolor{blue}{(3)}
\\
\hline
\end{tabular} 
\end{table}
\end{center}

\section{Is the observed $\sigma$ meson related with the chiral restoration?}

Is the pole observed in the $\pi$-$\pi$ phase shift really 
the $\sigma$ as the quantum fluctuation of the
order parameter of the chiral transition? 
A change of  the environment will lead to
that of the modes coupled to
 the order  parameter.
Thus production of the $\sigma$-meson in nuclear medium should be
useful for exploring the existence of the $\sigma$ and the possible
 restoration of chiral symmetry at finite density
\cite{tit,kuni95,nagahiro02,chiral02}.

One may ask what good observables is to see the softening in the 
$\sigma$ channel in nuclear medium.
A particle might loose its identity when put in a medium,
by various process like
$\sigma\leftrightarrow 2\pi$, 
$\sigma\leftrightarrow$  p-h, $\pi+$p-h, $\Delta$-h, $\pi+\Delta$-h
...\, 
Thus one needs to  calculate the strength function
of the hadron channel in the media\cite{ptp85,CH98}.

The surprise was,
such an enhancement had been seen by an Experiment by 
CHAOS collaboration\cite{chaos} at $T=0$ but at $\rho_B\not=0$
 by the reaction
${\rm A}(\pi^{+}, \pi^{+}\pi^{\pm}){\rm A}'$,
where the atomic number A runs from 2 to  208;
we remark that the experiment was   motivated  by other purpose
\cite{motivation} but not by exploring partial 
restoration of chiral symmetry in a nuclear medium.
Clearer information on the strength function in a
 nuclear medium may be provided with electro-magnetic probes;
 the reaction A($\gamma, \pi^0\pi^0$)A' performed by 
TAPS collaboration\cite{taps} clearly shows a softening of the cross section,
i.e., a downward shift of the cross section as a function of the invariant energy
of 2 pions, as shown in Fig.4.
\begin{figure}
\includegraphics[scale=.45]{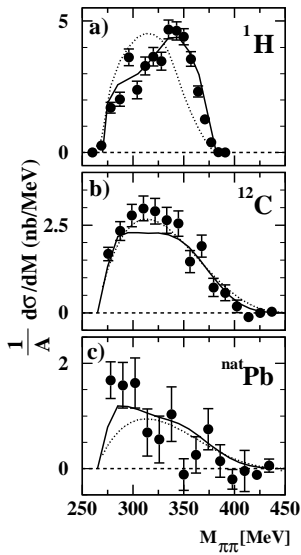}
\caption{
Photo-2$\pi$ cross section on nuclear targets\cite{taps}.
The incident energy belongs to the region 400 to 600 MeV.} 
\end{figure}

\begin{figure}
\includegraphics[scale=.45]{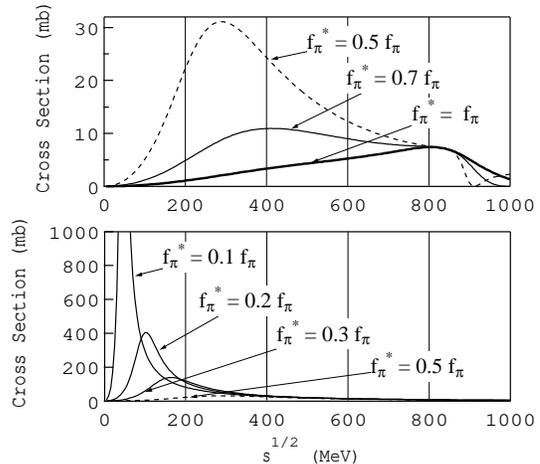}
\caption{
The  $T$ matrix in the $N/D$ method. 
      The in-medium $\pi$-$\pi$ cross sections in $I$=$J$=0 channel.
The upper (lower) panel shows the case of small (large) restoration
 corresponding to 
$0.5 f_\pi < f_\pi^* < f_\pi$ ($0.1 f_\pi < f_\pi^* < 0.5 f_\pi$).
Taken from \cite{yokokawa}.
\label{yokokawa1}
}
\end{figure}

\begin{figure}
\includegraphics[scale=.35]{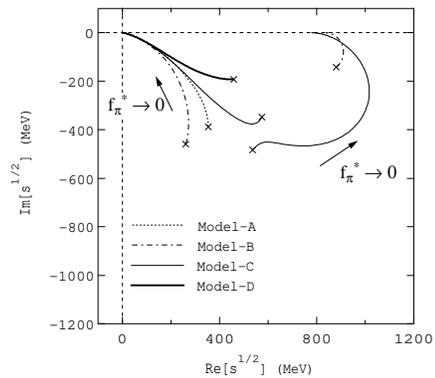}%
\caption{
 The movement of the poles in 
$I$=$J$=0  channel along with the  the decrease of $f_{\pi}^*$.
The crosses  are the pole positions in the vacuum.
Taken from \cite{yokokawa}.
\label{yokokawa2}}
\end{figure}

A calculation of the strength function in the $\sigma$ channel in 
a nuclear medium with  $\rho_B\not=0$
was made by Hatsuda et al \cite{HKS} based on a linear
$\sigma$ model; they showed that  an enhancement of the strength
function near 2$m_{\pi}$ threshold as observed in \cite{chaos, taps}
might be accounted for in terms of  a partial restoration of chiral 
symmetry in heavy nuclei.
Interestingly enough,
Jido et al \cite{jhk} showed that a similar enhancement can be obtained even in 
the nonlinear realization of chiral symmetry;
the key ingredient is  the wave-function renormalization of the pion field in the medium.
Subsequently, the $N/D$ method \cite{ND} a la Igi-Hikasa\cite{igi}is applied 
to examine the strength function in the scalar and vector channels 
by Yokokawa et al\cite{yokokawa}.
They found a softening of the spectral function both in 
the $\sigma$ and the $\rho$ meson channels being associated with partial restoration 
of chiral symmetry in the nuclear and hadron media:
 Fig.\ref{yokokawa1} shows the softening of the spectral 
function in the $\sigma$ 
channel;
following \cite{jhk}, the medium effect of chiral properties was
taken into account by the density-dependent pion decay constant
$f_{\pi}^{\ast}(\rho)$ which is closely related with the wave-function
renormalization mentioned above.
In Fig.\ref{yokokawa2}, shown is the softening of the $\sigma$ meson pole in the
  2nd Riemann sheet\cite{yokokawa}.

\section{Summary}
 Although there have been accumulation of 
experimental evidence of the
$\sigma$ pole with a low mass  500 to 600 MeV
in the pi-pi scattering matrix,
there are serious controversies on the nature of the $\sigma$ meson: 
 Naive quark model is in trouble for explaining
such a low-mass state in the $^3P_0$ state;
it may be a four-quark or  a collective q-$\bar{\rm q}$ or $\pi$-$\pi$
molecular state. The $\sigma$ as a collective q-$\bar{\rm q}$
state is identified as  the quantum fluctuation of the
 order parameter of the chiral transition.
The existence of such a collective mode in the scalar channel
can account for some  phenomena in hadron physics
which otherwise remain mysterious.

 To explore the nature of the $\sigma$ meson, 
trying to create a $\sigma$-mesonic mode
in  a nuclear medium may be helpful;
a peculiar enhancement of 
 the spectral function in the $\sigma$ channel 
in the lowering energy side  may be observed
along with a partial restoration of chiral symmetry in the medium.
 Such an enhancement might have  been already observed 
 in some experiments.

Recently, possible  $N_c$-dependence of the nature of the $\sigma$ meson 
has been noticed by some authors\cite{schafer,pelaez};
T. Schaefer showed that 
at $N_c=3$ the low mass $\sigma $ exists which is described as a linear combination
 of q-$\bar {\rm q}$  and $({\rm q}\bar{\rm q})^2$.
However, for larger $N_c$, $m_{\sigma}$ goes up and the $\sigma$ becomes
 mainly composed of q-$\bar{\rm q}$.
The same problem is examined  by others
but somewhat different conclusions are deduced\cite{pelaez}.
In passing, we  remark that a lattice simulation with dynamical fermions shows
a low-mass $\sigma$ degenerate with the $\rho$ meson\cite{muroya03}.

%

%



\begin{acknowledgments}
The present report 
utilized the  works done in collaboration with T. Hatsuda, 
K. Hayashigaki,  D. Jido,  H. Shimizu and
K. Yokokawa, to whom the author is grateful.
This work is supported by the Grants-in-Aids of the Japanese
Ministry of Education, Science and Culture (No. 14540263). 
\end{acknowledgments}


\end{document}